\documentclass{article}
\usepackage{epsfig}
\usepackage{amssymb} 
\usepackage{amsmath} 
\usepackage{amsfonts} 
\usepackage{cite} 
\usepackage{psfrag}

\newcommand{\beq}[1]{
\begin{equation}\label{#1}}
\newcommand{\eeq}{\end{equation}}
\newcommand{\bea}[1]{
\begin{eqnarray}\label{#1}}
\newcommand{\eea}{\end{eqnarray}}
\psfrag{p}{$p$}
\psfrag{pi}{$\pi$}
\psfrag{1-xi}{$1-\xi$}
\psfrag{1+xi}{$1+\xi$}
\psfrag{b}{${\bf b}$}
\psfrag{xib/(1+xi)}{$ \frac{\xi}{1+\xi} {\bf b}$}
\psfrag{xi/(1-xi)b}{$\frac{\xi}{1-\xi} {\bf b}$}
\psfrag{u}{$u$}
\psfrag{d}{$d$}
\psfrag{1/Q}{$\frac{1}{Q}$}
\psfrag{b}{${\bf b}$}
  
\psfrag{pp}{$p'$}
\psfrag{q}{$q$}
\psfrag{qp}{$q'$}
\psfrag{pip}{$\pi^+$}
\psfrag{pim}{$\pi^-$}
\psfrag{pr}{$P$}
\psfrag{apr}{$\bar P$}
\psfrag{g}{$\gamma$}
\psfrag{gs}{$\gamma^*$}
\psfrag{u}{$u$}
\psfrag{db}{$\bar d$}
\psfrag{ep}{$e^+$}
\psfrag{em}{$e^-$}
\psfrag{d}{$d$} 
\psfrag{a}{$a$}
\psfrag{TH}{$T_H$}
\psfrag{DA}{$DA$}
\psfrag{TDA}{$TDA$}

\begin{document}

\begin{titlepage}

\begin{center}

{\LARGE \bf
A QCD analysis of $\bar p N \to \gamma^*\pi $ and $ \bar p N 
\to \gamma^*\gamma$\\
Where is the pion in the proton ?}

\vspace{1cm}

{\sc B.~Pire}${}^{1}$,
{\sc L.~Szymanowski}${}^{2,3}$ 
\\[0.5cm]
\vspace*{0.1cm} ${}^1${\it
CPhT, {\'E}cole Polytechnique, F-91128 Palaiseau, France\footnote{
  Unit{\'e} mixte C7644 du CNRS.} \\[0.2cm]
\vspace*{0.1cm} ${}^2$  {\it
 So{\l}tan Institute for Nuclear Studies,
Ho\.za 69,\\ 00-681 Warsaw, Poland
                       } \\[0.2cm]
\vspace*{0.1cm} ${}^3$ {\it
Universit\'e  de Li\`ege,  B4000  Li\`ege, Belgium  } \\[1.0cm]
}

\end{center}
\vskip2cm

We   study   the   scaling  regime   of  nucleon   -
anti-nucleon  annihilation into a  deeply virtual  photon and  a photon or meson,  
$\bar p N \to \gamma ^* \pi $, $\bar p N \to \gamma ^* \gamma $, in the forward direction.
 The leading twist amplitude factorizes into 
an antiproton distribution amplitude,
a short-distance matrix element  and a long-distance dominated 
transition distribution amplitude (TDA)  which describes the nucleon to
meson or photon transition. 
 The  impact representation of this TDA maps out the transverse locations of 
 the small size core and the meson or photon cloud inside the proton.
\vskip1cm
%\end{center}

\vspace*{1cm}

\end{titlepage}

\section{A new factorization}
The understanding  of the hadronic structure needs appropriate tools to be manufactured\cite{AHP}.
It recently appeared that a fruitful approach could be accessed through exclusive hard quasi forward scattering, 
the prototype reaction being deep virtual Compton scattering in the forward region. We have generalized \cite{PS2,PS3} this 
analysis to the reactions 
$$\bar p N \to \gamma^*\pi ~~~~~~~~~~~~~~~ \bar p N 
\to \gamma^*\gamma$$
which will be accessible at future intense antiproton facilities\cite{PAX}. Our arguments for
the factorization of the short distance hard subprocess from the usual distribution amplitude and 
a new transition distribution amplitude, defined below, are a succession of logical steps generalizing
the factorization proof \cite{CFS} of deep exclusive meson electroproduction on a meson $\gamma^*
M_1 \to M_2 M_3$ in the forward direction, 
to its time reversed \cite{TCS} $M_2 M_1 \to \gamma^* M_3$, to 
meson-meson annihilation  $M_2 M_1 \to \gamma^* \gamma$ with the meson-photon analogy proven by the studies 
of the photon structure functions. The ultimate generalization from the meson case to the baryon case, 
implying three quark exchanges is advocated to be safe on the basis of the QCD analysis of baryon form factors.

We thus propose to write the $\bar p N \to \gamma^*\pi$ amplitude  as 
\begin{equation}
{\cal M} (Q^2,  \xi, t)= \int dx dy \phi(y_i,Q^2)
T_{H}(x_i, y_{i}, Q^2) T(x_{i}, \xi, t, Q^2)\;,
\label{amp}
\end{equation}
where $\phi(y_i,Q^2)$ is the antiproton distribution amplitude,
$T_{H}$ the hard scattering amplitude, calculated in the colinear approximation and $T(x_{i}, \xi, t, Q^2)$
the new TDAs.
 
% \vspace{1cm}
 % \vspace{1cm}

%%%%%%%%%%%%%%%%%%%%%  FIG. 1    %%%%%%%%%%%%%%%%%%%%%%%%%%%%%%%%%%%
\begin{figure}[h]
\begin{center}
{\includegraphics[width=6cm]{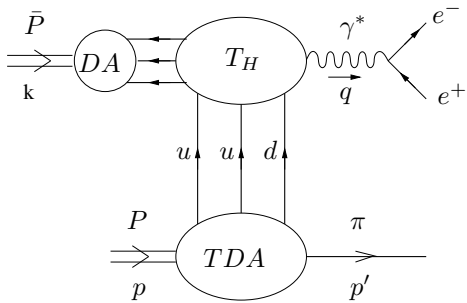}}
\caption{The factorization of the annihilation process $\bar p \;p\to \gamma^*
\,\pi$ into the antiproton distribution amplitude $(DA)$, 
the hard subprocess amplitude $(T_H)$ and a baryon $\to$ meson transition
distribution amplitude $(TDA)$.}
\label{angular}
\end{center}
\end{figure}
%%%%%%%%%%%%%%%%%%%%%%%%%%%%%%%%%%%%%%%%%%%%%%%%%%%%%%%%%%%%%%%

\section{Transition Distribution Amplitudes}
To define the  transition distribution
amplitudes  from a nucleon to a pseudoscalar meson, we introduce light-cone coordinates $v^\pm = (v^0 \pm
v^3) /\sqrt{2}$ and  transverse components $v_T = (v^1,  v^2)$ for any
four-vector $v$.   The skewedness variable  $\xi = -\Delta^+  /2P^+$ with $\Delta = p'-p$ and
$P=(p+p')/2$  describes  the loss  of  plus-momentum  of the  incident
hadron in the proton $\to$ meson transition. We parametrize the quark momenta as 
shown on Fig. 1. The fractions of + 
momenta are labelled $x_{1}$, $x_{2}$ and $x_{3}$, 
and their supports are within $[-1+\xi, 1+\xi]$.  
Momentum conservation implies  :
$\sum_{i}  x_{i} = 2 \xi \, .$
The fields with positive momentum fractions, $x_i\ge 0$, describe creation of 
quarks, whereas those with  negative momentum fractions, $x_i\le 0$, the 
absorption of antiquarks.
The eight leading twist TDAs for the $p \to \pi^0$ (which can be 
expressed in terms of eight 
independent helicity amplitudes for $p \to uud\;\pi$ transition) then reads :
\bea{TDA}
%\begin{eqnarray}
%\label{TDA}
 &&  4  \langle     \pi^{0}(p')|\, 
\epsilon^{ijk}u^{i}_{\alpha}(z_1\,n) u^{j}_{\beta}(z_2\,n)d^{k}_{\gamma}(z_3\,n)
\,|p(p,s) \rangle \Big|_{z^+=0,\,  z_T=0}  
\\ \nonumber
&&= -\frac{f_N}{2f_\pi}\left[ V^{0}_{1} (\hat P C)_{\alpha\beta}(B)_{\gamma}  +
A^{0}_{1} (\hat P\gamma^5 C)_{\alpha\beta}(\gamma^5 B)_{\gamma} -
3\,T^{0}_{1} ( P^\nu i\sigma_{\mu\nu} C)_{\alpha\beta}(\gamma^\mu B)_{\gamma} \right] 
\nonumber \\
&&+ V^{0}_{2} 
 (\hat P C)_{\alpha\beta}(\hat \Delta_{T} B)_{\gamma} +
A^{0}_{2}(\hat P \gamma^5 C)_{\alpha\beta}(\hat \Delta_{T}\gamma^5 B)_{\gamma}
+ T^{0}_{2} (\Delta_{T}^\mu P^\nu \sigma_{\mu\nu} C)_{\alpha\beta}(B)_{\gamma}
\nonumber \\
&&+  T^{0}_{3} ( P^\nu \sigma_{\mu\nu} C)_{\alpha\beta}(\sigma^{\mu\rho}
\Delta_{T}^\rho B)_{\gamma} + \frac{T_{4}^0}{M } (\Delta_{T}^\mu P^\nu 
\sigma_{\mu\nu} C)_{\alpha\beta}(\hat \Delta_{T}¥B)_{\gamma}\;, \nonumber
%\end{eqnarray}
\eea
where $\sigma^{\mu\nu}= i/2[\gamma^\mu, \gamma^\nu]$, $C$ is the charge 
conjugation matrix 
and $B$ the nucleon spinor. 
$f_\pi$ is the pion decay constant ( $f_\pi = 93$ MeV) and $f_N$ is the
constant which determines the value of the nucleon wave function at the
origin, and which has been estimated through QCD sum rules to be of
order $5.3\cdot 10^{-3}$ GeV$^2$ .
Each TDA is then  Fourier transformed to get the usual representation in terms of the 
momentum fractions, through the relation
\begin{equation}
F (z_{i}P\cdot n) = \int\limits^{1+\xi}_{-1+\xi} d^3x 
\delta (x_{1}+ x_{2}+ x_{3} -2\xi) e^{-iPn\Sigma x_{i}z_{i}} \, F(x_{i},\xi)
\end{equation}
where $F$ stands for $V_{i}, A_{i}, T_{i}$ and $\int d^3x \equiv 
\int dx_1dx_2dx_3 \delta(2\xi - x_1 -x_2-x_3)$.
The first three 
terms in (\ref{TDA}) are the only ones surviving the forward limit
$\Delta_T \to 0$.
The constants in front of these three  
terms  have been chosen
in reference to  the soft pion ($\xi \to 1$) limit results :
\bea{softp}
&&V^0_1(x_1,x_2,x_3) \to  
\frac{1}{2}\left( \phi_N(x_1,x_2,x_3)
+\phi_N(x_2,x_1,x_3) \right) 
\nonumber \\
&&A^0_1(x_1,x_2,x_3) \to  \frac{1}{2}\left( 
\phi_N(x_1,x_2,x_3)
-\phi_N(x_2,x_1,x_3) \right)
 \\
&&T^0_1(x_1,x_2,x_3) \to 
\frac{1}{2}\left( \phi_N(x_1,x_3,x_2)
+\phi_N(x_2,x_3,x_1) \right) \nonumber \;, 
\eea
where 
$\phi_N(x_1,x_2,x_3)$ is the standard  leading twist DA.

\section{Impact Parameter Picture}
As in the case of generalized parton distributions \cite{GPD} 
and distribution amplitudes \cite{PS1}
the simultaneous presence of two transverse 
scales $Q^2$ and $-t$, allows through a Fourier transform
to map the impact parameter dependence of the scattering amplitude. 
In the case under study, the $t-$ dependence
of the $N \to \pi$ transition distribution amplitude allows in 
its ERBL region (namely, when all $x_{i}> 0$) a transverse scan of 
the location of the small sized 
(of the order of $1/Q$ ) hard core made of 
three quarks when a pion carries the rest of the momentum of the nucleon.
This may be phrased alternatively as detecting 
the transverse mean position of a pion inside the proton, 
when the proton 
state is of the "next to leading Fock " order, namely $ |\, qqq \,\pi >$. 
This is shown on Fig. 2. The other regions have slightly different 
interpretations.

%%%%%%%%%%%%%%%%%%  FIG. 2 %%%%%%%%%%%%%%%%%%%%%%%%%%
\begin{figure}
\begin{center}
{\includegraphics[width=7cm]{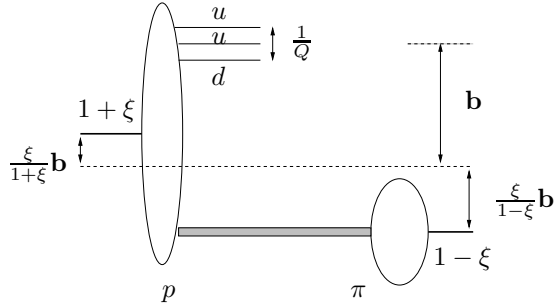}}
\caption{Impact parameter space representation of the $p\,\to\,\pi^0$ TDA 
in the ERBL region $x_i>0$.}
\label{angular}
\end{center}
\end{figure}
%%%%%%%%%%%%%%%%%%%     FIGURE 2          %%%%%%%%%%%%%%%%%%%%%%%%%%%%

The study of different TDAs such as the ones for $p\to \pi^{0}$ and $n\to \pi^{-}$, related to the $\bar p p$ 
and $\bar p n$ reactions, may shed light on the $ |\, uud \,\pi^0 >$ versus $ |\, udd \,\pi^{+} >$ components of the proton.
 
 \section{Conclusion}
 The formalism developed for the proton antiproton exclusive annihilation  may  as well be used
 for related channels such as backward virtual Compton scattering 
or backward electroproduction of a meson, 
 where data exist for moderate values of $Q^2$. These spacelike analogs of the processes discussed here share the same
 virtues and their  studies should allow a first look at the internal structure of the  $ |\, qqq \,\pi >$ states inside the 
 nucleon. Mesonic channels may also be studied as $ \gamma^*\gamma \to \pi \pi $ , $\gamma^*\gamma \to \pi \rho $
 or $\gamma^*\gamma \to \rho \rho$ in the near forward region. The TDAs are then not much different from the 
 mesonic GPDs. 
 
\vskip.1in
\noindent
Work of L.Sz. is supported by the Polish Grant 1 P03B 028 28. 
He is a Visiting Fellow of the FNRS (Belgium).

\end{document}